\documentclass[conference]{IEEEtran}
\IEEEoverridecommandlockouts
\usepackage{cite}
\usepackage{amsmath,amssymb,amsfonts}
\usepackage{algorithm}
\usepackage{algpseudocode}
\usepackage{graphicx}
\usepackage{textcomp}
\usepackage{booktabs} 
\usepackage{xcolor}
\usepackage{tikz}
\usepackage{amsthm}
\usepackage{enumitem}

\usepackage{hyperref}

\usepackage{xspace}
\usepackage[colorinlistoftodos, color=blue!30!white %   ,shadow
]{todonotes}                                        
\newtheorem{remark}{Remark}
\newtheorem{theorem}{Theorem}

\newtheorem{proposition}{Proposition}
\def\BibTeX{{\rm B\kern-.05em{\sc i\kern-.025em b}\kern-.08em
    T\kern-.1667em\lower.7ex\hbox{E}\kern-.125emX}}

\begin{document}

\title{An Algorithm for Computing the Capacity of Symmetrized KL Information for Discrete Channels}

\author{\IEEEauthorblockN{Haobo Chen }
\IEEEauthorblockA{\textit{Department of ECE} \\
\textit{University of Florida}\\
Gainesville, US \\
haobo.chen@ufl.edu}
\and
\IEEEauthorblockN{Gholamali Aminian}
\IEEEauthorblockA{
\textit{The Alan Turing Institute}\\
London, UK  \\ gaminian@turing.ac.uk
}
\and
\IEEEauthorblockN{Yuheng Bu}
\IEEEauthorblockA{\textit{Department of ECE} \\
\textit{University of Florida}\\
Gainesville, US  \\
buyuheng@ufl.edu}
}

\maketitle

\begin{abstract}
% Traditional KL divergence's asymmetry is addressed through a symmetrized version, offering a balanced measure for distribution similarity.
Symmetrized Kullback-Leibler (KL) information (\(I_{\mathrm{SKL}}\)),
which symmetrizes the traditional mutual information by integrating Lautum information, has been shown as a critical quantity in communication~\cite{aminian2015capacity} and learning theory~\cite{aminian2023information}.
This paper considers the problem of computing the capacity in terms of \(I_{\mathrm{SKL}}\) for a fixed discrete channel. Such a maximization problem is reformulated into a discrete quadratic optimization with a simplex constraint. One major challenge here is the non-concavity of Lautum information, which complicates the optimization problem. Our method involves symmetrizing the  KL divergence matrix and applying iterative updates to ensure a non-decreasing update while maintaining a valid probability distribution.
We validate our algorithm on Binary symmetric Channels and Binomial Channels, demonstrating its consistency with theoretical values. Additionally, we explore its application in machine learning through the Gibbs channel, showcasing the effectiveness of our algorithm in finding the worst-case data distributions. 
% This work advances the understanding of channel capacity and input distribution impacts, offering significant insights into information theory and practical applications.
\end{abstract}


\section{Introduction}

% The study of information theory is central to understanding communication channels, data distributions, and their capacities. 
Kullback-Leibler (KL) divergence, as a widely adopted measure in information theory, quantifies the difference between two probability distributions. Specifically, the KL divergence between two probability measures \(P\) and \(Q\) over a space \(\mathcal{X}\), where \(P\) is absolutely continuous with respect to \(Q\), is defined as
\begin{align}
D(P\|Q) = \int_{\mathcal{X}} P(x) \log \frac{P(x)}{Q(x)} \, dx.
\end{align}
% In general, the KL divergence is not symmetric, i.e., .
However, KL divergence is not symmetric in general, i.e., \(D(P\|Q) \neq D(Q\|P)\), which limits its applicability in some contexts. To address this issue, the symmetrized KL divergence, or Jeffrey's divergence~\cite{jaynes1957information}, has been introduced to provide a symmetric measure of distribution similarity, which is defined as
\begin{align}
D_{\mathrm{SKL}}(P\|Q) \triangleq D(P\|Q) + D(Q\|P).
\end{align}
This symmetrized version of KL divergence exhibits inherent symmetry and simplicity, making it applicable in various scenarios as a measure of similarity for distributions, e.g. see~\cite{pereira2024asymptotics,aminian2021exact,aminian2023information,aminian2015capacity}.
% \bu{Please specify the scenarios or applications in more detail.}
% \chen{Professor, could you help me by adding a list of papers that discuss the application of symmetrized KL information?}
% \bu{We can mention its role in characterizing the generalization error~\cite{aminian2021exact,aminian2023information} and the usage as the upper bound for channel capacity~\cite{aminian2015capacity}. }

The mutual information between two random variables \(X\) and \(Y\) is defined as the KL divergence between their joint distribution and the product of their marginal distributions:
\begin{align}
I(X;Y) \triangleq D(P_{XY} \| P_X P_Y).
\end{align}
Similarly, the symmetrized KL information (\(I_{\mathrm{SKL}}\)) between \(X\) and \(Y\) is given by:
\begin{align*}
I_{\mathrm{SKL}}(X;Y) \triangleq D_{\mathrm{SKL}}(P_{XY} \| P_X P_Y) = I(X;Y) + L(X;Y),
\end{align*}
where \(L(X;Y) \triangleq  D(P_X P_Y\|P_{XY}) \) denotes the Lautum information~\cite{palomar2008lautum}.

This paper explores the capacity problem with symmetrized KL information, i.e., maximizing \(I_{\mathrm{SKL}}\) over a fixed channel, which can be applied to bound channel capacity~\cite{aminian2015capacity} and understand the worst-case generalization error as in~\cite{aminian2023information}.

% We present a novel approach to compute this capacity using a discrete quadratic optimization problem. Our method leverages iterative refinement within simplex constraints, ensuring convergence to an optimal solution.

Maximizing \(I_{\mathrm{SKL}}\) poses significant challenges due to the non-concave nature of the Lautum information with respect to the input distribution. Therefore, traditional methods for maximizing mutual information, e.g., Arimoto-Blahut algorithm \cite{Blahut1972ComputationOC}, which rely on the concavity of the problem, are not directly applicable here. 

% To address the non-concavity issue,
% we present a novel approach to compute this capacity by reformulating the problem as a discrete quadratic optimization with a simplex constraint. Our method leverages iterative refinement within simple constraints.

To address the non-concavity issue,
we reformulate the problem into a quadratic matrix optimization over probability simplex. Specifically, in Section~\ref{sec:matrix}, we derive a matrix representation of \(I_{\mathrm{SKL}}\) in the discrete case. We further symmetrize this matrix representation and apply iterative updates with element-wise normalization, ensuring adherence to the simplex constraints. This iterative algorithm, detailed in Section~\ref{sec:algorithm}, is designed to handle the non-concave nature of the problem, guaranteeing a monotonic update in terms of $I_{\mathrm{SKL}}$. 

% show how this can be transformed into a quadratic optimization problem with simplex constraints.

We apply our algorithm to various channel models, including the Binary Symmetric Channel (BSC) and Binomial Channel, validating the proposed algorithm with empirical data. The results presented in Section~\ref{sec:exp} show a strong alignment between the calculated capacities and theoretical values, demonstrating the effectiveness of our approach. Additionally, we explore the implications of our method in a machine-learning context through the Gibbs channel, showcasing its versatility in finding the worst-case data distributions as in Section~\ref{sec:Gibbs}.

\section{Preliminaries}

% The Kullback-Leibler (KL) divergence is widely adopted in information theory and statistics to quantify the difference between two distributions. 
% To address the lack of symmetry, the symmetrized KL information, also known as Jeffrey's divergence\cite{jaynes1957information}, is defined as:
% \begin{align}
% D_{\mathrm{SKL}}(P\|Q) \triangleq D(P\|Q) + D(Q\|P).
% \end{align}
\subsection{Information measures}

We note that the symmetrized KL divergence also belongs to the \(f\)-divergence family~\cite{sason2016f}, with $f(x) = x\log x-\log x$.
As an \( f \)-divergence, symmetrized KL divergence has the following variational representation,
\begin{align}\label{eq: rep SKL}
    D_f(P\|Q) = \sup_{h} \mathbb{E}_P[h(X)] - \mathbb{E}_Q[f^*(h(X))],
\end{align}
where \( f^*(\cdot) \) denotes the Legendre transform of function $f$.

Total variation (TV) is another important measure in information theory and statistics that quantifies the difference between two probability distributions. It is defined as:
\begin{align}
\mathrm{TV}(P, Q) \triangleq \sup_{A} |P(A) - Q(A)|,
\end{align}
where the supremum is taken over all measurable sets \(A\). For discrete distributions, the total variation can be expressed as:
\begin{align}
\mathrm{TV}(P, Q) = \frac{1}{2} \sum_{x \in \mathcal{X}} |P(x) - Q(x)|.
\end{align}

% Total variation provides a measure of the largest possible difference between the probabilities assigned to the same event by two distributions. We will use this notation as the convergence check in our algorithm later.

\subsection{Capacity}
The concept of channel capacity is crucial in information theory, as it captures the fundamental limits of communication.
% which is defined as the maximum rate at which information can be reliably transmitted over the channel. 
It is well-known that channel capacity can be characterized by maximizing mutual information for the channel \(P_{Y|X}\):
\begin{align}
C = \max_{P_X} I(X;Y),
\end{align}
where the maximization is over all possible input distributions defined over \(\mathcal{X}\).

Analogous to channel capacity, the capacity using symmetrized KL information for a fixed channel \(P_{Y|X}\) is
\begin{align}\label{equ:C_SKL}
    C_{\mathrm{SKL}} = \max_{P_X \in \mathcal{P}} I_{\mathrm{SKL}}(X;Y),
\end{align}
where the maximization is over some convex set of input distributions \(\mathcal{P}\). Such a capacity has different interpretations in various fields.

$C_{\mathrm{SKL}}$ plays a significant role in bounding channel capacity. It has been shown \cite{aminian2015capacity} that for any channel 
% \(W\), where \(x\) represents the input signal and \(y\) represents the output, 
the capacity \(C\) can be upper bounded by:
\begin{align}
C \leq C_{\mathrm{SKL}}.
% \max_{P_X} I_{\mathrm{SKL}}(X;Y).
\end{align}
% This expression indicates that a channel's maximum achievable information transfer rate (capacity) can be understood through the lens of symmetrized KL divergence. 
In \cite{aminian2015capacity}, the authors demonstrated the similarity between the capacities defined by mutual information and symmetrized KL information in various channels, including the Binary Symmetric Channel and point-to-point Gaussian channels. Their study also shows that the capacity of memoryless Poisson channels, based on symmetrized KL divergence, is computable. Furthermore, numerical results suggest that this upper bound is effective for channels in molecular communication with small capacities. By rigorously analyzing the point-to-point Gaussian channel and the Poisson channel, they demonstrated the broad applicability of this upper bound across different types of communication channels.

% \chen{I have added Gholamali’s result here.}
% \bu{The discussion here can be more specific, e.g., describing Gholamali's result on the Poisson channel by converting standard capacity with SKL.}

Besides its role in bounding channel capacity, $C_{\mathrm{SKL}}$ can be interpreted as the worst-case generalization error for the Gibbs
channel within the context of learning theory. It is shown in~\cite{aminian2021exact,aminian2023information} that the generalization error can be characterized using the $I_{\mathrm{SKL}}$ between the input training data and the learned weights. Therefore, the capacity-achieving input distribution for $I_{\mathrm{SKL}}$ can be interpreted as the worst-case data input distribution. This concept aligns with findings in the literature that generalization behavior is highly data-dependent, and a single learning algorithm does not work for all kinds of training data. Thus, examining the capacity of symmetrized KL information provides significant insights into understanding the data-dependent nature of generalization in learning theory.

% which aligns with findings in the literature on information-theoretic limits of learning, showing that certain data distributions (e.g., images) generalize well, while others do not.
% For example, it can represent the largest expected error in \cite{aminian2021exact} within the context of learning theory, achieved by the worst-case input distribution \(P_X^*\). 
% Thus, examining the capacity of symmetrized KL information yields significant insights into the mapping relationship between \( X \) and \( Y \) and its implications for learning theory. 

% It also facilitates a nuanced understanding of model capacity within the context of information theory and signal processing for channels with fixed characteristics.

\subsection{Related Works}
% This subsection discusses the challenges of optimizing symmetrized KL information over the input distribution. 

% \todoa{Do you want to represent this as proposition?} 

\textbf{Channel Capacity Computation:} In the context of maximizing mutual information, it is crucial to design an effective capacity computation algorithm for various scenarios. For discrete channels, most algorithms rely on iterative numerical methods. These include the famous Arimoto-Blahut algorithm \cite{Blahut1972ComputationOC, Yu_2010}, linearly constrained optimization approaches \cite{https://doi.org/10.1002/nav.3800030109, Li2023LearningCC}, and simulation-based numerical algorithms \cite{1661831, 965976}. 
These methods often require an exact characterization of the discrete channel. 
% often succeed in discrete cases due to their ability to analytically calculate mutual information by its definition. 
Therefore, these conventional algorithms cannot be easily extended to general continuous channels, where only pairs of training samples for channel input and output are available.
To resolve such a limitation, a sample-based mutual information estimator is essential for handling such capacity estimation problems.

% they do not include a specific sample-based mutual information estimator, which is essential for handling general continuous channels. This limitation leads to underperformance in continuous input channels.

% A method to navigate this complexity will be proposed.

% \bu{We need a paragraph to explain why we should care about the concavity of the problem. The standard problem is easy as it is concave, but it is not the case for SKL.}
% \chen{Revised for gap between the optimization problem in Mutual information and symmetrized KL information}

% This innovation offers empirical insights into the optimal input distribution for neural network models.

\textbf{Channel Capacity Estimation:}  Channel capacity estimation can be viewed as two fundamental tasks: (a) estimating the mutual information using samples of the channel input and output and (b) maximizing this mutual information with respect to the channel input distribution. Traditional methods for the former task include binning \cite{hacine2018binning}, non-parametric kernel estimation \cite{bi2018high, gretton2003kernel}, and Gaussian distribution approximation \cite{hulle2005edgeworth}. However, these traditional methods lack scalability and struggle with large sample sizes and dimensions, particularly for high-dimensional data. The latter task can be addressed using gradient descent methods for differentiable mutual information estimators.
With advancements in deep learning, significant progress has been made in mutual information estimation. Recent work combines variational methods with neural networks to construct neural network-based estimators of mutual information \cite{belghazi2021mine,fritschek2019deep,tsur2022neural,10075335,aharoni2020capacity,Mirkarimi_2021,sreekumar2022neural,9449919}. 
These methods have also been applied to channel capacity estimation, leveraging high-dimensional encoders with neural network\cite{letizia2021discriminative, belghazi2018mine, song2019understanding}.

% Although estimating symmetrized KL divergence is somewhat tractable with fixed channels,

% \textbf{$\mathbf{I_{\mathrm{SKL}}}$ Estimation:} For specific channels, the symmetrized KL information has closed-form expression, which can be easily estimated, allowing us to determine which input distributions can maximize symmetrized KL information.  Previous studies have utilized mean estimation for certain Gaussian distributions to explore the relationship between generalized data distributions \cite{aminian2021exact}. Additionally,  \cite{yao2024symmetric} employed Stein’s method along with the Central Limit Theorem to address the intractable computation of symmetrized KL divergence. These approaches facilitate the observation of convergence rates between zero-mean normalized sums of independent random variables and the standard normal distribution. However, this estimation lacks concavity with respect to the input distribution, precluding the use of established mutual information capacity analysis properties and related analytical theorems. This necessitates alternative strategies for the effective maximization of symmetrized KL information.

% \chen{Add the previous challenge part in here}

\section{Challenges}
% A reassessment reveals that the Lautum information is not concave with respect to the marginal input, which makes the problem a non-concave optimization. A formal discussion of the non-concavity issue is provided in Appendix~\ref{app:non-concave}. 

Both discrete algorithm and neural network-based continuous methods for computing channel capacity rely on the concavity of mutual information over input distribution for a fixed channel. In \cite{palomar2008lautum}, it was claimed that Lautum information is concave for a fixed \( P_{Y|X} \). 

However, upon re-examining the proof steps, we show that Lautum information is not concave. 
% \bu{Let's make it a formal theorem}

\begin{theorem}
For a fixed channel \( P_{Y|X} \), Lautum information $L(X;Y)$ is not concave with respect to the input distribution $P_X$.
\end{theorem}

The non-concavity of Lautum information is due to the fact that the chain rule for mutual information, which is crucial for demonstrating concavity, does not hold for Lautum information~\cite{palomar2008lautum}. The detailed proof of this result is provided in Appendix~\ref{app:non-concave}. Therefore, the $C_{\mathrm{SKL}}$ problem, which integrates mutual information with Lautum Information, is not concave, presenting significant challenges.

% This re-evaluation transforms the original problem of finding \( C_{\mathrm{SKL}} \) from a concave to a non-concave one. 

We also note that variational methods that maximize mutual information using the variational representation of $f$-divergence cannot be directly applied to symmetrized KL information. In particular, the Legendre transform of $f(x) = x\log x-\log x$ is \( f^*(\cdot) = \exp(t + \text{LambertW}(\exp(1 - t)) - 1) \), which relies on the Lambert W function~\cite{corless1996lambertw}, presents computational challenges for iterative updates using neural networks~\cite{belghazi2021mine}. 

\section{Proposed Method}
In this section, we tackle the challenges associated with the capacity problem of symmetrized KL information for discrete channels. We propose a novel approach, transforming the problem into a quadratic optimization problem and introducing an iterative algorithm to ensure that the objective function is always non-decreasing.

\subsection{Matrix Representation of \(I_{\mathrm{SKL}}\) in the Discrete Case}\label{sec:matrix}

To simplify the capacity problem  in~\eqref{equ:C_SKL}, we consider an alternative expression for $I_{\mathrm{SKL}}$ as shown in the following proposition.
% a fixed conditional distribution \( P_{Y|X} \) within a discrete framework. 

\begin{proposition}\label{prop:SKL}
For fixed channel \( P_{Y|X} \), \( I_{\mathrm{SKL}} \) can be expressed equivalently as\:
\begin{align}
I_{\mathrm{SKL}}(X; Y) \!= \!\sum_{x,\tilde{x}} P_X(x) P_X(\tilde{x}) D\left( P_{Y|X\!=x} \| P_{Y|X\!=\tilde{x}} \right),
\end{align}
where \( \tilde{X} \) denotes an independent copy of \( X \), sharing the same distribution \( P_{X} \).
\end{proposition}
% \bu{Move the proof to the appendix}
The detailed proof can be found in Appendix \ref{app:SKL}.
Using Proposition~\ref{prop:SKL}, we can reformulate the capacity problem in~\eqref{equ:C_SKL} into a quadratic matrix optimization problem subject to simplex constraints. To see this, we begin by defining the vector \(\boldsymbol{X} \in \Delta^{d-1}\), where $\Delta^{d-1}$ denotes the probability simplex with dimension $d-1$. If the size of the space $|\mathcal{X}|=d$, then each element \( \boldsymbol{X}_i \) in vector $\boldsymbol{X}$ represents the probability \( P_X(x_i) \) for $x_i \in \mathcal{X}$. 
% The dimension of \(\boldsymbol{X}\) is \( d \times 1 \).
Next, we define the matrix \(\boldsymbol{C}\), which is an \( d \times d \) non-negative matrix where each entry \( C_{ij} \) represents the KL divergence \( D(P_{Y|X=x_i} \| P_{Y|X=x_j}) \). 

% The dimension of \(\boldsymbol{C}\) is \( d \times d \).

With these notations, by Proposition~\ref{prop:SKL}, \( I_{\mathrm{SKL}} \) can be written in matrix form,
\begin{equation}
    I_{\mathrm{SKL}}(X; Y) = \boldsymbol{X}^\top \boldsymbol{C} \boldsymbol{X}.
\end{equation}
To ensure that \(\boldsymbol{X} \in \Delta^{d-1}\), which represents a valid probability distribution, we impose the following simplex constraints: \(\boldsymbol{X}^\top \boldsymbol{1} = 1\) and \(\boldsymbol{X} \geq \boldsymbol{0}\), where \(\boldsymbol{1}\) is an \(d\)-dimensional all-one vector.

Combining all these components, we obtain the following equivalent quadratic matrix optimization problem:
\begin{align}\label{equ:matrix_capacity}
\max_{\boldsymbol{X}\in \mathbb{R}} \quad & \boldsymbol{X}^\top \boldsymbol{C} \boldsymbol{X} \nonumber \\
\text{subject to} \quad & \boldsymbol{X}^\top \boldsymbol{1} = 1, \\
& \boldsymbol{X} \geq \boldsymbol{0}. \nonumber
\end{align}
% where \(\boldsymbol{X}\) is an \( d \)-dimensional vector representing the probability distribution \( P_X \), \(\boldsymbol{C}\) is an \( d \times d \) matrix with entries \( C_{ij} = D(P(Y|X=i) \| P(Y|X=j)) \), and \(\boldsymbol{1}\) is an \( d \)-dimensional vector of ones.

% This reformulation shows how the original expression for \( I_{\mathrm{SKL}} \) can be transformed into a quadratic matrix optimization problem with a simplex constraint. 

This reformulation leverages the properties of the KL divergence and probability distributions, providing a clear path for optimization. Noteworthy observations include:
\begin{enumerate}[label=(\alph*)]
\item The matrix \(\boldsymbol{C}\) is not symmetric and contains non-negative elements, since the KL divergence \(D(P_{Y|X=x_i} \| P_{Y|X=x_j})\) are asymmetric and non-negative;
\item As \(D(P_{Y|X=x} \| P_{Y|X=x})=0\), all diagonal elements of \(\boldsymbol{C}\) are zero; 
\item The matrix \(\boldsymbol{C}\) is not definite, as the capacity problem for \( I_{\mathrm{SKL}} \) is not concave nor convex.
\end{enumerate}
Therefore, this quadratic optimization problem is not directly solvable, necessitating the exploration of specific problem structures to design an appropriate algorithm.

% \bu{Let's discuss if we want to state it as a theorem. Also, define those notions religiously, e.g., what are the dimensions of these vectors and matrix? }
% \chen{Add more detail about the dimension of the formulation here.}

\subsection{Max-SKL Algorithm }
\label{sec:algorithm}
% \todoa{It would be better to change Max-SKL to Max-SKL.}
In this section, we provide a novel iterative algorithm for solving the optimization problem in~\eqref{equ:matrix_capacity}, which ensures that the objective function is always non-decreasing. 

First, we symmetrize the KL divergence matrix \(\boldsymbol{C}\) in~\eqref{equ:matrix_capacity}
% Let \(\boldsymbol{C}\) be a \(d \times d\) matrix with entries \(C_{ij} = D(P(Y|X=i) \| P(Y|X=j))\). 
by averaging it with its transpose:
\begin{align}\label{equ:sym}
    \boldsymbol{C}_{\text{sym}} = \frac{1}{2}(\boldsymbol{C} + \boldsymbol{C}^\top).
\end{align}
It is straightforward to verify that $\boldsymbol{X}^\top \boldsymbol{C} \boldsymbol{X} = \boldsymbol{X}^\top \boldsymbol{C}_{\text{sym}} \boldsymbol{X}$, and we will only work with symmetric $\boldsymbol{C}_{\text{sym}}$ in the following.

Here, we notice that this non-definite simplex-constrained quadratic problem is related to the variational representation of the eigen-problem~\cite{horn2012matrix}. In particular, for any symmetric matrix $\boldsymbol{A}$, the maximum of the following optimization problem with a norm constraint
\begin{align}\label{equ:matrix_eigen}
\max_{\boldsymbol{X}} \quad & \boldsymbol{X}^\top \boldsymbol{A} \boldsymbol{X}  \\
\text{subject to} \quad & \|\boldsymbol{X}\|_2 = 1, \nonumber
% \\
% & \boldsymbol{X} \geq \boldsymbol{0}, \nonumber
\end{align}
is the largest eigenvalue of $\boldsymbol{A}$, which is achieved when $\boldsymbol{X}$ equals to the corresponding eigenvector. Furthermore, as $\boldsymbol{C}_{\text{sym}}$ is a non-negative symmetric matrix, Perron-Frobenius theorem \cite[Theorem
8.2.2]{horn2012matrix} states that its dominant eigenvector is element-wise non-negative, and the constraint $\boldsymbol{X} \geq \boldsymbol{0}$ is automatically satisfied. 

Therefore, the optimization problem in~\eqref{equ:matrix_capacity} can be viewed as replacing the norm constraint in~\eqref{equ:matrix_eigen} with a simplex constraint. We anticipate that existing algorithms for finding the largest eigenvector can be adapted to solve~\eqref{equ:matrix_capacity}. We then introduce our algorithm for updating probability distributions within the simplex constraint, which is motivated by the power iteration method. 

% , so eigenvalue-based optimization methods, such as those derived from Perron-Frobenius theorem \cite{horn2012matrix}, cannot be directly applied.

% Thus, the task evolves into optimizing a non-negative simplex-constrained quadratic problem. 

Power iteration is a well-known algorithm used to find the dominant eigenvector of a matrix through iterative updates and normalization \cite{GoluVanl96}. Specifically, the power iteration algorithm proceeds as follows. Given a matrix \(\boldsymbol{A}\), the algorithm starts with an initial vector \(\boldsymbol{x}_0\) and iteratively updates it with the following rule
% \chen{Introduce our algorithm using power iteration here.}
\begin{align}
    \boldsymbol{x}_{k+1} = \frac{\boldsymbol{A} \boldsymbol{x}_k}{\|\boldsymbol{A} \boldsymbol{x}_k\|}.
\end{align}
% where \(\|\cdot\|\) denotes the norm (typically the Euclidean norm). 
This process will converge when \(\boldsymbol{x}_k\) reaches the dominant eigenvector of \(\boldsymbol{A}\). However, this normalization by the vector norm only guarantees the norm constraint in~\eqref{equ:matrix_eigen} but cannot ensure that the sum of all elements in \(\boldsymbol{x}_k\) is one.

% \chen{I have item the algorithm here}
% \todoa{If you have space, itemize the following discussion.}: 
Our algorithm leverages a similar iterative process but is specifically designed for the simplex constraint. 
In the proposed iterative update, the following operations are performed:
\begin{enumerate}[label=(\alph*)]
    \item Compute \(\boldsymbol{C}_{\text{sym}} \boldsymbol{X}_k\), which is the product of the symmetrized matrix and the current probability distribution vector.
    \item Update the probability distribution by multiplying the current distribution \(\boldsymbol{X}_k\) \emph{element-wise} with \(\boldsymbol{C}_{\text{sym}} \boldsymbol{X}_k\).
    \item Normalize the updated probability distribution to ensure that the probabilities sum to 1.
\end{enumerate}
In particular, starting with an initial \(\boldsymbol{X}_0 \in \Delta^{d-1}\), we consider the element-wise update rules 
\begin{align}
    [\boldsymbol{X}_{k+1}]_i = \frac{[\boldsymbol{X}_k]_i \cdot [(\boldsymbol{C}_{\text{sym}} \boldsymbol{X}_k)]_i}{\boldsymbol{X}_k^\top \boldsymbol{C}_{\text{sym}} \boldsymbol{X}_k}, \quad i = 1, \ldots, d,
\end{align}
where $[\cdot]_i$ denotes the $i$-th element of the vector. As both $\boldsymbol{X}_k$ and $\boldsymbol{C}_{\text{sym}}$ are non-negative, the resulting solution will satisfy $\boldsymbol{X} \geq \boldsymbol{0}$ automatically. This approach leverages the structure of the symmetrized matrix $\boldsymbol{C}_{\text{sym}}$ to find a solution that satisfies the non-negativity and simplex constraints.

% Despite these differences, both algorithms involve multiplying a matrix by a vector and require normalization after each iteration to maintain certain vector properties (unit norm for power iteration, sum to 1 for the proposed algorithm).

% The approach is supported by a lemma from \cite{Scheuer1959}, ensuring monotonicity and convergence:

In addition, the following lemma from \cite{Scheuer1959} ensures the monotonicity and convergence of the proposed iterative process, which guarantees that \(\boldsymbol{X}_{k+1}^\top \boldsymbol{C}_{\text{sym}} \boldsymbol{X}_{k+1} \geq \boldsymbol{X}_k^\top \boldsymbol{C}_{\text{sym}} \boldsymbol{X}_k\).
% , which follows from the monotonicity of the iterative process.

 % \(d\) elements with relative viabilities \(C_{ij} \geq 0\), forming
\begin{proposition}[{\cite[Section 3]{Scheuer1959}}]
Consider a symmetric \(d \times d\) matrix \(\boldsymbol{C}_{\text{sym}} = (C_{ij})\) with non-negative elements $C_{ij} \geq 0$. Let \(x^k_i\) be the probability of the \(i\)-th element \((i = 1, 2, \ldots, d)\) of $\boldsymbol{X}_k$ in iteration \(k\), such that \(\sum_{i=1}^{d} x_i = 1\) and \(0 < x_i < 1\).

Let 
% \todoa{The current notation in this Lemma should be revised.}
\begin{align}
    I_{\mathrm{SKL}}^k &= \boldsymbol{X}_k^\top \boldsymbol{C}_{\text{sym}} \boldsymbol{X}_k= \sum_{i=1}^{d} \sum_{j=1}^{d} C_{ij} x_i^k x_j^k,\\
     I_{\mathrm{SKL}}^{k+1} &= \boldsymbol{X}_{k+1}^\top \boldsymbol{C}_{\text{sym}} \boldsymbol{X}_{k+1}.
\end{align}

% The frequencies $\left[\boldsymbol{X}_{k+1}\right]_i$ of the elements in the next state are given by:

% \begin{align}
%     &I_{\mathrm{SKL}}^k \left[\boldsymbol{X}_{k+1}\right]_i = [\boldsymbol{X}_k]_i \cdot [(\boldsymbol{C}_{\text{sym}} \boldsymbol{X}_k)]_i \\
%     &=x_i^k \sum_{j=1}^{d} C_{ij} x_j^k \quad \text{for all} \quad i = 1, 2, \ldots, d.
% \end{align}

% Let 
% \begin{align}
   
% \end{align}

% The value \(I_{\mathrm{SKL}}^{k+1}\) is monotonic with respect to the new frequencies $\left[\boldsymbol{X}_{k+1}\right]_i$, satisfying the inequality:
Then we have the following inequality,
\begin{align}
    I_{\mathrm{SKL}}^{k+1} - I_{\mathrm{SKL}}^{k}  \geq 0.
\end{align}
\end{proposition}

% \bu{The description and the notation used in this theorem are still hard to digest. Why not use the matrix notation? What do you want to emphasize here?}

This Proposition validates the convergence and correctness of the update rules used in our algorithm, ensuring that the objective is always non-deceasing under the given constraints. Our algorithm will terminate when the total variation between $\boldsymbol{X}_k$ and $\boldsymbol{X}_{k+1}$ is smaller than a threshold $\epsilon >0$.
The overall description of the proposed Max-SKL algorithm can be found in Algorithm~\ref{alg:RefinementPx}.

\begin{remark}[Comparison with Power iteration algorithm]
Our algorithm shares similarities with the power iteration method but also has distinct differences. The power iteration algorithm aims to find the dominant eigenvector of a matrix, utilizing direct matrix-vector multiplication and normalization using vector norm. However, it does not guarantee that the sum of the elements of the vector remains one after each iteration. In contrast, our proposed algorithm ensures that the sum of the probability distribution vector elements is one by using a normalization based on the element-wise product.
\end{remark}

% In summary, our algorithm is not identical to the power iteration method, it shares iterative refinement and convergence characteristics. It also incorporates a normalization mechanism suited for probability distributions, making it analogous but distinct from traditional power iteration methods.

% \bu{There is a huge gap between the optimization problem we defined and the proposed algorithm. We cannot just formulate the problem and provide a solution directly, you need to mention the intuition and the flow behind the algorithm. We can discuss the related eigen problem and Perron theorem, and these results motivate us to use this algorithm. }

% \bu{We only state monotonicity in the lemma; the description of the algorithm should be outside.}

% \bu{Let's come up with a better name for the algorithm Max-SKL algorithm }
% \todoa{Could we replace $\| P(X)_{k+1} - P(X)_k \|_1 < \epsilon$ with $TV()\leq \epsilon$?}
\begin{algorithm}[t]
\caption{Max-SKL algorithm}\label{alg:RefinementPx}
\begin{algorithmic}[1]
\State $\boldsymbol{C} \gets$ Compute the KL divergence matrix with all elements $C_{ij} = D(P_{Y|X=x_i} \| P_{Y|X=x_j})$
\State $\boldsymbol{C} \gets \frac{1}{2}(\boldsymbol{C} + \boldsymbol{C}^\top)$ 
\State $\boldsymbol{X}_0 \gets$ Initialize the probability distribution over $\mathcal{X}$
\For{$k \gets 0$ \textbf{to} $maxIter - 1$}
    \State $CP \gets \boldsymbol{C} \cdot \boldsymbol{X}_k$ \Comment{Matrix-vector product}
    \State $\boldsymbol{X}_{k}^{'} \gets \boldsymbol{X}_k \ast CP$ \Comment{Element-wise product}
    \State $\boldsymbol{X}_{k+1} \gets \boldsymbol{X}_{k}^{'} / (\boldsymbol{X}_k^\top \cdot CP)$ 
    % \If{$\| P(X)_{k+1} - P(X)_k \|_1 < \epsilon$}
    \If{$\mathrm{TV}(\boldsymbol{X}_{k+1}, \boldsymbol{X}_k) \leq \epsilon$}
        \State \textbf{break} \Comment{Convergence check}
    \EndIf
\EndFor
\State \Return $\boldsymbol{X}_{k+1}$, $\boldsymbol{X}^T \boldsymbol{C}_{\text{sym}} \boldsymbol{X}$
\end{algorithmic}
\end{algorithm}

% \bu{Can we have some simpler experiments that are not relevant to Gibbs? It is the best that 
% 1. we know the analytical solution of $P_X^*$
% 2. We can compare the numerical solution to the true $P_X^*$
% 3. The optimal solution $P_X^*$ is different from the acid of the mutual information. 
% }

\section{Experiments with Discrete Channels}
\label{sec:exp}
To demonstrate the efficacy of our Algorithm \ref{alg:RefinementPx}, we conduct various experiments under different settings. 
% These experiments serve to validate the theoretical models and demonstrate the algorithm's practical applicability. 
These experiments showcase the algorithm's ability to handle different types of channels and input distributions, providing a reliable tool for optimizing symmetrized KL information in practical applications.  
% The detailed results and convergence behaviors further support the algorithm's practical applicability and theoretical soundness.
\subsection{Binary Symmetric Channel}
To validate the proposed Algorithm \ref{alg:RefinementPx} for finding the symmetric KL capacity, we conduct an experiment for the Binary Symmetric Channel (BSC). As shown in~\cite{aminian2015capacity}, the theoretical capacity of a BSC with crossover probability of $p \in [0,1]$ is given by:
\begin{equation}\label{equ:C_BSC}
    C_{\mathrm{SKL}}(p) = -\log_2 \left( \sqrt{p(1 - p)} \right) - h(p),
\end{equation}
where \( h(p) \) is the binary entropy function.
% defined as:
% \begin{equation}
% h(p) \triangleq -p \log_2(p) - (1 - p) \log_2(1 - p).
% \end{equation}
The theoretical $C_{\mathrm{SKL}}$ reaches its maximum at a uniform input probability distribution. Our goal is to validate our algorithm by computing the capacity and comparing it with theoretical values.

\begin{figure}[t]
    \centering
    \includegraphics[width=0.5\textwidth]{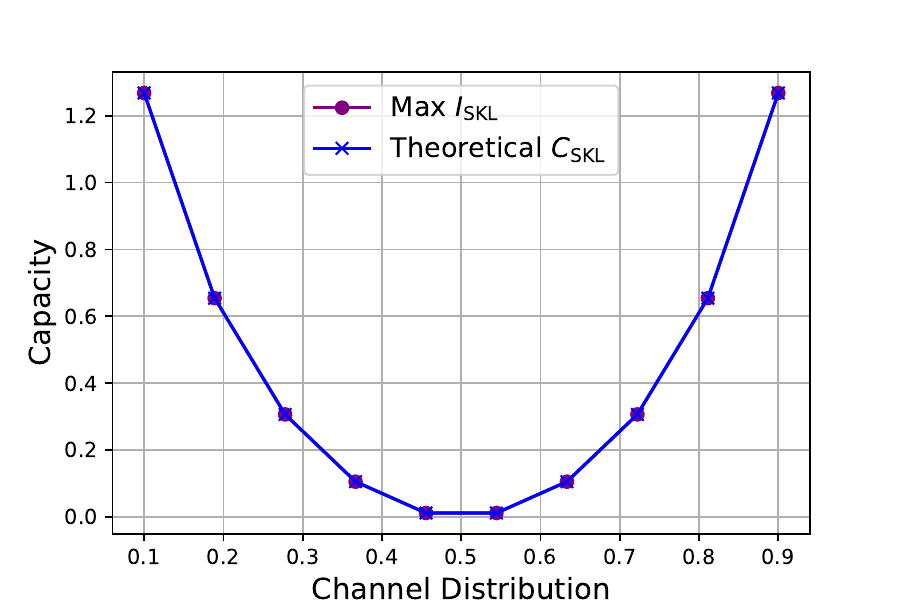}
    \caption{Comparison of theoretical values and calculated symmetric KL capacities using Max-SKL algorithm for the BSC. The experiment varies the channel distribution \( p \) to validate that our algorithm can accurately compute the theoretical \( I_{\mathrm{SKL}} \).}\label{fig:C_BSC}
\end{figure}

In Figure~\ref{fig:C_BSC}, We compare the calculated capacities by Algorithm \ref{alg:RefinementPx} with the theoretical values \( C_{\mathrm{SKL}}(p) \) using~\eqref{equ:C_BSC} for different crossover probabilities \( p \) ranging from 0.1 to 0.9. 
% The theoretical capacity for each \( p \) was computed using~\eqref{equ:C_BSC} for \( C_{\mathrm{SKL}}(p) \). 
% The calculated capacity was obtained using Algorithm~\ref{alg:RefinementPx} and computing \( X^T \boldsymbol{C}_{\text{sym}} X \).
 % compares the theoretical capacities with the calculated capacities, 
Figure~\ref{fig:C_BSC} shows an excellent agreement between the two, which confirms that our algorithm effectively maximizes the symmetrized KL information.
% \bu{Add the description of the new baselines.}
% \todoa{It would be better to import figures with EPS format.}

% Our algorithm-optimized input achieves the symmetric KL capacity 
The capacity-achieving input distribution (caid) of $C_{\mathrm{SKL}}$ identified by our algorithm is a uniform distribution, which corresponds to the theoretical caid for $C_{\mathrm{SKL}}$ and is the same as the caid of traditional capacity with mutual information. To further validate our algorithm, we conduct additional experiments on Binary Asymmetric Channel (BAC) in Appendix~\ref{App:BAC} to show that the capacity-achieving input distributions can be different for different capacities. 

% it can achieve the objective value for different channels.

\subsection{Binomial Channel}
We consider a binomial channel~\cite{aminian2015capacity} where the input of the channel \(X\) is a continuous value in the range \(\mathcal{X}=[0, 1]\) and the output \(Y\) is a discrete variable taking values in \(\{0, 1, \ldots, n\}\). The relationship between \(X\) and \(Y\) is described by the binomial distribution:
\begin{equation}\label{equ:bino}
    P(Y=y|X=x) = \binom{n}{y} x^y (1-x)^{n-y}.
\end{equation}
For our experiments, we quantize the input space $\mathcal{X}$ ranging from 0.1 to 0.9 with increments of 0.1 so that it is a discrete channel. The channel $P(Y|X)$ is calculated using the binomial distribution in ~\eqref{equ:bino} with $n=10$.
% \begin{itemize}
% \item we set channel parameter \(n = 10\);
% \item ;
%     % \item \textbf{Output \(Y\)}: Values from 0 to 10.
% \item .
% \end{itemize}

% \subsection{Mutual Information and \(I_{\mathrm{SKL}}\)}
We consider two baseline algorithms in the comparison:
\begin{itemize}
    \item Arimoto-Blahut algorithm, which is designed to maximize the mutual information over fixed channel;
    \item Power iteration designed for solving eigen problem.
\end{itemize}
In addition, we also compare the Algorithm~\ref{alg:RefinementPx} with its counterpart without the symmetrizing step in~\eqref{equ:sym}, denoted as  $\text{Max-SKL}$ and $\text{Max-SKL-wos}$, respectively. 

% \textbf{\(\boldsymbol{I_{\mathrm{SKL}}}\)}: Can be simplified as the covariance \(\text{cov}(Y, \log(X(1-X)))\) and optimized using the 

% The initial probabilities \(P(X)\) are uniformly distributed for the element-wise optimization and randomly initialized for the power iteration.
% \subsubsection{Optimal Input Distribution for Mutual Information}

Using the Arimoto-Blahut algorithm, caid  for mutual information is given by:
% \vspace{-1em}
\begin{equation*}
    P(X) = [0.36, 0.00, 0.00, 0.05, 0.18, 0.05, 0.00, 0.00, 0.36].
\end{equation*}
% \vspace{-1em}
% \vspace{-1em} % Adjust this value as needed to reduce the space
% \begin{figure}[H]
%     \centering
%     \includegraphics[width=0.5\textwidth]{pictures/blahut_arimoto_convergence (1).pdf}
%     \caption{Convergence of Arimoto-Blahut Algorithm for Mutual Information.}
%     \label{fig:ab_convergence}
% \end{figure}
% \subsubsection{Optimal Input Distribution for \(I_{\mathrm{SKL}}\)}
Using the Max-SKL algorithm, the caid for $I_{\mathrm{SKL}}$ is:
% \vspace{-2em}
\begin{equation*}
P(X) = [0.5, 0, 0, 0, 0, 0, 0, 0, 0.5]
\end{equation*}

\begin{figure}[t!]
    \centering
    \includegraphics[width=0.5\textwidth]{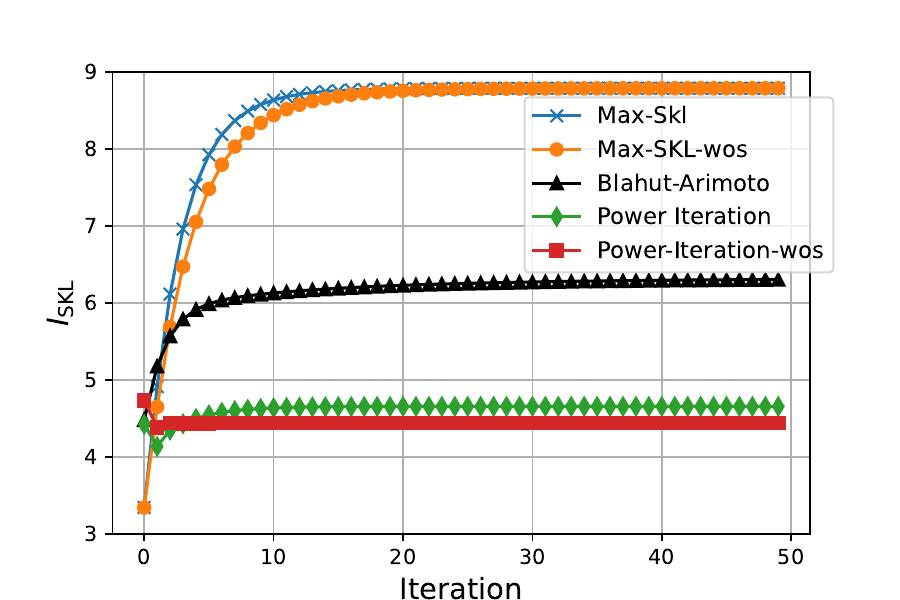}
    \caption{Convergence of Max-SKL and Power Iteration, and comparison with the result of the Blahut-Arimoto algorithm to calculate \(I_{\mathrm{SKL}}\). Our algorithm shows successful convergence in the binomial channel, with the Max-SKL using a symmetrizing step demonstrating the best performance.}
    \vspace{-1em}    \label{fig:optimization_convergence}
\end{figure}

% \chen{I replace the Binary Asymmetric Channel with binomial channel, do you think the fig 2 and fig3 should in same plot?}

\begin{figure}[t]
    \centering
    \includegraphics[width=0.5\textwidth]{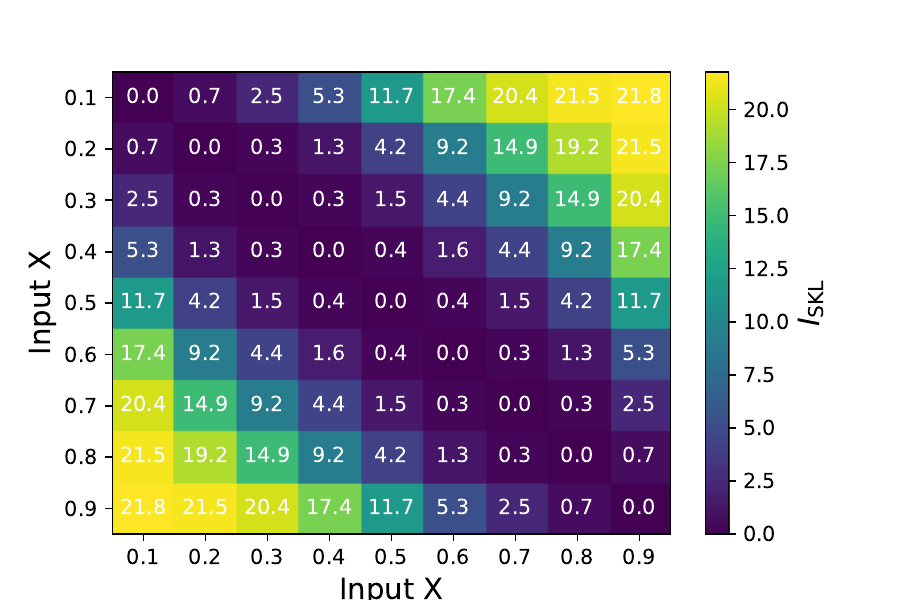}
    \caption{Matrix of KL Divergence \(C_{sym}\) for \(n=10\). The entries at 0.1 and 0.9 are the largest, leading to a concentrated distribution at these points to maximize \(X^T CX\).}
    \label{fig:kl_divergence_n10}
\end{figure}

% \subsection{Comparison of Methods}
The convergence of \(I_{\mathrm{SKL}}\) over iterations are illustrated in Figure~\ref{fig:optimization_convergence}.
Our algorithm successfully converges in the binomial channel, and the $\text{Max-SKL}$ demonstrated the best performance.
We also apply the Blahut-Arimoto algorithm to maximize the mutual information and use its caid to calculate the \(I_{\mathrm{SKL}}\) for comparison with our results. We found that the Lautum information term plays a significant role in maximizing \(I_{\mathrm{SKL}}\). Additionally, the caids for mutual information and \(I_{\mathrm{SKL}}\) are quite different. The caid for \(I(X;Y)\) is more spread out, whereas for \(I_{\mathrm{SKL}}\), the distribution is concentrated at specific points.

To better understand the caid for  \(I_{\mathrm{SKL}}\), we visualize the KL divergence matrix \(C_{sym}\) in Figure~\ref{fig:kl_divergence_n10}. 
When \(n=10\), we observe that the entries at 0.1 and 0.9 in \(C_{sym}\) are the largest. Therefore, to maximize \(X^T CX\), the distribution should concentrate on these two boundary points. However, when \(n=100\) (see Figure~\ref{fig:kl_divergence_n100} in Appendix \ref{Bino}), the caid spreads out more evenly across other points.
% \todoa{Why Bold text?}

\section{Experiments with the Gibbs Channel}\label{sec:Gibbs}
% \todoa{Do you want to cite Raginsky paper?}, as introduced by Raginsky \cite{NIPS2017_ad71c82b}, 

In this section, we demonstrate the effectiveness of our algorithm for a specific channel, where $I_{\mathrm{SKL}}$ can be interpreted as the generalization error of a supervised learning problem,
highlighting the significance of the capacity problem in learning theory. Before delving into the experimental setup, we first provide the context for the Gibbs channel.

\subsection{Dataset and Loss Function}
We consider a supervised learning problem with training dataset \( S = \{(Z_i)\}_{i=1}^n \), where each data \(Z_i = (X_i, Y_i) \in \mathcal{S}\) is i.i.d. generated from the data distribution \(P_{Z}\). The performance of different machine learning models $W\in \mathcal{W}$ is measured by a loss function \(\ell\colon \mathcal{W} \times \mathcal{Z} \to \mathbb{R}^+\). Here, we consider the following Mean Squared Error (MSE) loss function
\begin{equation}
    \ell(w, z) =\|y - \hat{y}\|_2^2 = \|y - x^\top w\|_2^2,
\end{equation}
with $\hat{y} = x^\top w$, i.e., we adopt a linear model.
The goal of supervised learning is to find a $w$ that minimizes the following population risk
\begin{equation}
L_P(w, P_Z) = \mathbb{E}_{P_Z}[\|Y - X^\top w\|_2^2].
\end{equation}
However, as data distribution $P_Z$ is unknown, we can only minimize the empirical risk, which can be written as:
\[
L_E(w, S) = \frac{1}{n} \sum_{i=1}^{n} (Y_i - X_i^\top w)^2 = \frac{1}{n} \|\mathbf{Y} - \mathbf{X}w\|_2^2,
\]
where $\mathbf{X},\mathbf{Y}$ stacks the $n$ i.i.d samples in vector form. 

\subsection{Gibbs Channel}
Any learning algorithm can be viewed as a channel that randomly maps the training dataset $S$ onto a model $W$ according to the probability transition matrix $P_{W|S}$.
Thus, the generalization error quantifying the degree of over-fitting can be written as
\begin{equation}\label{Eq: expected GE}
\mathrm{gen}(P_{W|S},P_S)\triangleq\mathbb{E}_{P_{W,S}}[ L_P(W, P_Z)-L_E(W,S)].
\end{equation}
% where the expectation is taken over the joint distribution $P_{W,S} =  P_{W|S}\otimes P_S$. 

% We consider the channel as the posterior and the likelihood with the loss function. 
% As evidenced in \cite{zhang2006information}, the solution to this information risk minimization task is encapsulated in the \( (\gamma, \pi(w), L_E(w, S)) \)-Gibbs algorithm, delineated as:
Here, we consider the Gibbs distribution as a channel, which is well-studied in Bayesian learning because it provides a principled way to incorporate prior knowledge and quantify uncertainty \cite{wenzel2020good,adlam2020cold}. 
Specifically, the \( (\gamma, \pi(w), L_E(w, S)) \)-Gibbs algorithm is given by
\begin{equation}
P^{\gamma}_{W|S}(w|S) = \frac{\pi(w)e^{-\gamma L_E(w, S)}}{\int_{\mathcal{W}}\pi(w)e^{-\gamma L_E(w, S)} dw},
\end{equation}
where $\pi(w)$ is any prior distribution and $\gamma >0$ denotes inverse temperature. 

In addition, as shown in~\cite{aminian2021exact,aminian2023information},
the Gibbs channel has the nice  property that its generalization error equals the symmetrized KL information between the input training samples and the learned model 
\begin{equation}
    \mathrm{gen}(P^{\gamma}_{W|S},P_S) = \frac{I_{\mathrm{SKL}}(W;S)}{\gamma}.
\end{equation}
Therefore, for any fixed Gibbs channel \(P^{\gamma}_{W|S}\), our Max-SKL algorithm can be used to identify the worst-case data input that maximizes the generalization error.

% To investigate the Gibbs channel for the \(n\)th sample, we select a prior that is derived from the Gibbs posterior distribution trained on the previous \(n-1\) samples. For this purpose, an initial prior distribution \(\pi(w)\) with zero mean and variance \(\frac{1}{n}\) is used during the training of the Gibbs posterior. 

% Given the prior $\pi(w) \sim \mathcal{N} \left(0, \frac{\sigma}{n} \mathbf{I}_p \right)$, where $\sigma$ and $\gamma$ are chosen as 1 and the number of samples ($n$) respectively, and the MSE loss function setup, the Gibbs channel is Gaussian distribution:\todoa{Is the following $P_{W|S}$?}
% \todoa{Where do you define $\pi(w)$ in terms of $\gamma$ and then you set $\gamma=n$?}

% \begin{align}
% P_{W|S} \sim \mathcal{N} \left( (n \mathbf{I}_{d}+\mathbf{X}^T\mathbf{X})^{-1}\mathbf{X}^T \mathbf{Y}, \frac{1}{n} (n \mathbf{I}_{d}+\mathbf{X}^T\mathbf{X})^{-1} \right).
% \end{align}
% Here, the prior distribution is zero mean and its variance is set to \(\frac{1}{n}\), ensuring that the parameters used in the Gibbs posterior distribution align with the prior assumptions.

% \todoa{This sentence is not clear.}

% \bu{Please specify the choice of $\gamma$ and $\pi$ here, as $\sigma_{N}$ is not defined.}
% \chen{I set $\sigma$ is set to 1 and $\gamma$ is chosen as the sample number here. }
\subsection{Experimental Setup}
% \chen{I am trying to use visualization to illustrate the Gibbs experiment here.}
In the following, we provide two toy examples to show how our algorithm can be used to find the discrete worst-case data distribution for the Gibbs channel. Using the MSE loss function and Gaussian prior distribution $\pi(w) \sim \mathcal{N} \left(0, \frac{1}{n} \mathbf{I} \right)$, the Gibbs channel can be modeled as a Gaussian posterior distribution. For our setup, we set the parameter \(\gamma\) equal to the number of samples \(n\), ensuring that the posterior appropriately reflects the sample size. 

Since our Max-SKL algorithm is designed for discrete input distributions, we make the following simplification:
\begin{enumerate}
    \item We restrict the feature $X$ to contain only two binary features $X \in \{-1,1\}^2$ with a binary label $Y \in \{-1,1\}$;
    \item Instead of solving the worst case distribution $P_S$ for $n$ samples directly, where the alphabet size will grow exponentially with $n$, we focus on the case where the prior distribution $\pi_n(w)$ is pre-trained with $n$ samples generated from $P_{S_0}$, and our goal is to identify the worst case distribution $P_{S_1}$ for the next $n+1$-th sample. 
\end{enumerate}

Thus, in this toy example, we only need to find a distribution with an alphabet size of 8.

% To demonstrate the application of our algorithm, we have structured the following experimental setup. 

% The dataset \(Z = (X, Y)\) comprises data pairs \((x_i, y_i)\) generated independently and identically distributed (i.i.d.) from the distribution \(P_X\). Here, \(x_i = [x_{i1}, x_{i2}, \ldots, x_{id}]\) are vectors in \(\mathbb{R}^d\), and \(y_i\) denotes the corresponding scalar outcomes. The functional relationship between \(X\) and \(Y\) is defined by a mapping function \(f_w\).

In the subsequent sections, we will discuss two specific cases where the pre-trained prior $\pi_n(w)$ differs because it is trained using the samples generated from different $P_{S_0}$.

% Gibbs posterior derived from the previous \(n-1\) 
% We consider the following setting, where the Gibbs posterior distributions are updated with each new data point. Given the prior data matrices \(X_{\text{prior}}\) and outcomes \(Y_{\text{prior}}\), and the total number of samples \(n\), we compute the posterior distributions as follows:
% % \todoa{Could you pls clarify more? the current form is not clear.}
% \begin{align*}
% \text{Prior of the } &(n)^\text{th} \text{ sample} = \text{Posterior of the } (n-1)^\text{th} \text{ sample}, \nonumber\\
% &\pi_n(w) = \frac{\pi(w) e^{-\gamma L_{E}(w, s_{n-1})}}{\mathbb{E}_{\pi}\big[ e^{-\gamma L_{E}(W, s_{n-1})}\big]},\ \text{ for }\ \gamma > 0.
% \end{align*}

% % \subsection{Dataset and Relationship between \(X\) and \(Y\)}

% The dataset \( Z=(X,Y) \) consists of data pairs \((x_i, y_i)\) generated i.i.d. from the distribution \(P_X\). Here, \(x_i = [x_{i1}, x_{i2}, \ldots, x_{id}]\) are vectors in \(\mathbb{R}^d\), and \(y_i\) represents the corresponding scalar outcomes. The relationship between \(X\) and \(Y\) is defined by a function \( f_w \), mapping input features to outcomes.

\subsection{Case 1: Linearly Separable}

\subsubsection{Motivation and Setup}

In this case, the input features \(X\) and the target outcomes \(Y\) of $P_{S_0}$ are linearly separable, meaning that a single hyperplane can accurately classify the data points into their respective classes. This setup helps us understand the impact of the worst-case data distribution on a model that initially performs well.

We consider the joint distribution $P_{S_0}$ defined uniformly over the following four points,
% specific data pairs \((x_i, y_i)\) with the following probabilities:
\begin{align*}
P_{S_0}([1, 1], 1) = 0.25, & \quad  P_{S_0}(([1, -1], -1) =0.25, \\
 P_{S_0}([-1, 1], 1)=0.25, & \quad P_{S_0}([-1, -1], -1)=0.25.
\end{align*}
% For simplicity, we assume all data probabilities are 0.25.

\subsubsection{Pre-training Process}

Figure~\ref{Fig: LS_case} illustrates  \( P_{S_0} \) of four data points with coordinates \([1, 1]\), \([1, -1]\), \([-1, 1]\), and \([-1, -1]\):
\begin{itemize}
    \item Class 1 (\( y = 1 \)): \([1, 1]\), \([-1, 1]\)
    \item Class -1 (\( y = -1 \)): \([1, -1]\), \([-1, -1]\).
\end{itemize}
It can be seen that \( y = x_2 \), i.e., $w = [0,1]^\top$ effectively fits the data points generated from $P_{S_0}$ into their correct classes. 

This pre-training uses the $n=100$ samples generated from $P_{S_0}$, which gives us the following prior for the $n+1$-th sample
\begin{align}\label{equ:prior_n}
\pi_n(w) = \frac{\pi(w) e^{-n L_{E}(w, s_{n})}}{\mathbb{E}_{\pi}\big[ e^{-n L_{E}(W, s_{n})}\big]}.
\end{align}

% The model uses these points to generate the initial fitting plane.

\subsubsection{Worst-Case Distribution Impact}

\begin{figure}[t!]
    \centering
    \includegraphics[width=0.48\textwidth]{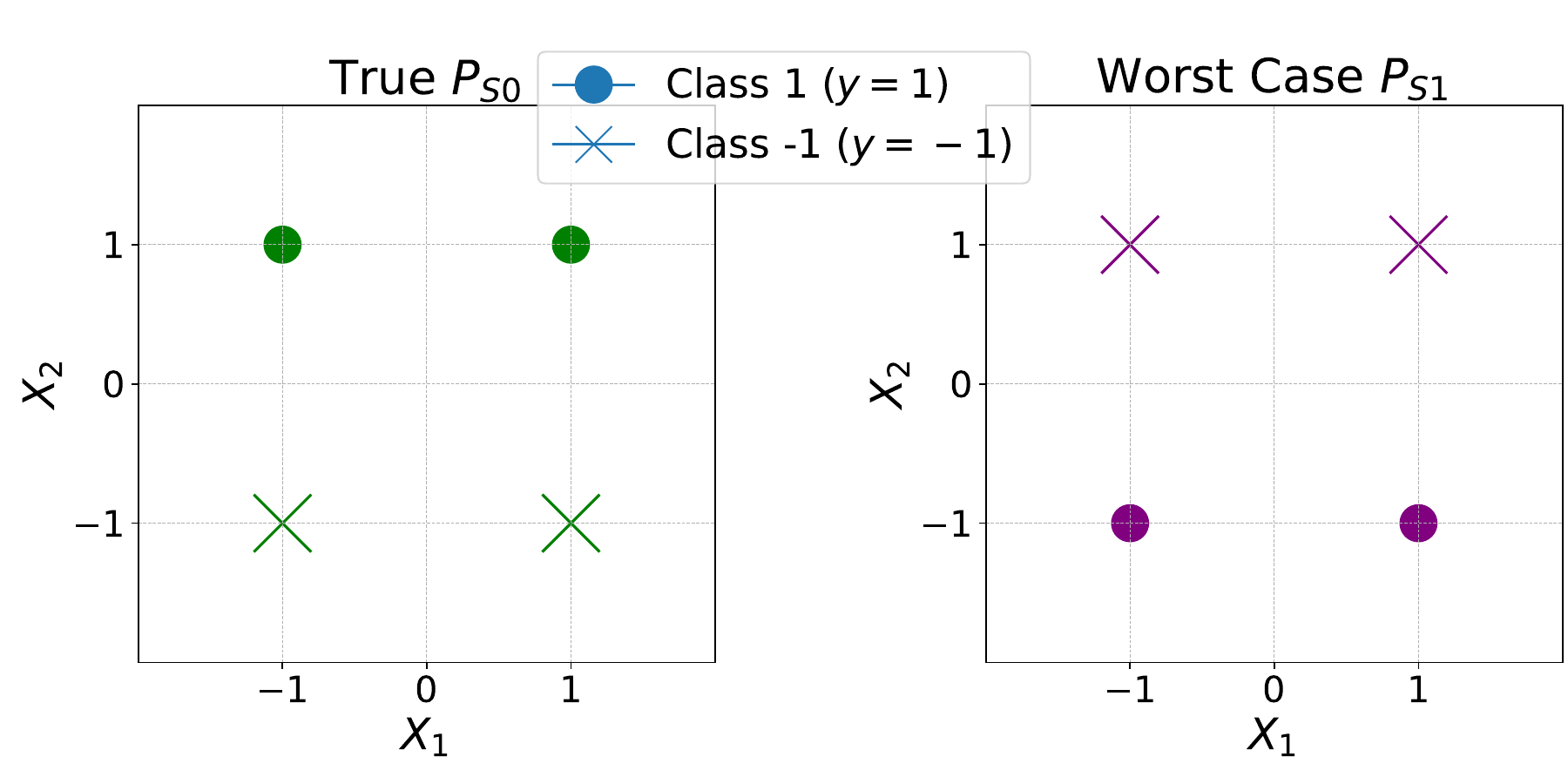}
    
    \caption{Linearly Separable Data Points (Case 1). The plots show the data points under two distributions: the initial distribution \( P_{S_0} \) and the worst-case distribution \( P_{S_1} \). In each plot, circles represent Class 1 (\( y = 1 \)) and crosses represent Class -1 (\( y = -1 \)). The initial model correctly fits the data points, but under the worst-case distribution, the fitting plane shifts, leading to misclassification.}
    \label{Fig: LS_case}
\end{figure}

Using Algorithm~\ref{alg:RefinementPx}, we identify the worst-case data distribution for the \( (n, \pi_n(w), L_E(w, Z_{n+1}))\)-Gibbs algorithm. Figure \ref{Fig: LS_case} visualizes the worst-case data distribution \( P_{S_1} \). While it is defined in the same coordinates as \( P_{S_0} \), the target labels have changed, resulting in altered class assignments. In \( P_{S_1} \):
\begin{itemize}
    \item Class 1 (\( y = 1 \)): \([1, -1]\), \([-1, 1]\)
    \item Class -1 (\( y = -1 \)): \([1, 1]\), \([-1, -1]\)
\end{itemize}
In Figure \ref{Fig: LS_case}, circles represent Class 1 (\( y = 1 \)) and crosses represent Class -1 (\( y = -1 \)). The worst-case distribution \( P_{S_1} \) alters the fitting plane to \( y = -x_2 \). This change results in the projection of the input data to the opposite class, thus degrading the model's performance. 

% The influence of the worst-case distribution is clearly shown in Fig.\ref{Fig: LS_case}.

% The series of plots \todoa{Which series of plots?} illustrates the progression of the fitting plane and the data point classifications from the true distribution \( P_{S_0} \) to the worst-case distributions \( P_{S_1} \) and \( P_{S2} \).

\subsection{Case 2: Linearly Non-separable}

\subsubsection{Motivation and Setup}

In this case, the input features \(X\) and the target outcomes \(Y\) are linearly non-separable, meaning no single hyperplane can correctly classify all the data points. This scenario is designed to understand the impact of the worst-case data distribution on a model that already faces challenges due to non-separability.

We consider the joint distribution \(P_{S_0}\) defined uniformly over the following four points:
\begin{align*}
P_{S_0}([1, 1], 1) = 0.25, & \quad P_{S_0}([1, -1], -1) = 0.25, \\
P_{S_0}([-1, 1], -1) = 0.25, & \quad P_{S_0}([-1, -1], 1) = 0.25.
\end{align*}

\subsubsection{Pre-training Process}

Figure~\ref{Fig: LNS_case} illustrates the distribution \(P_{S_0}\) for the data points with coordinates \([1, 1]\), \([1, -1]\), \([-1, 1]\), and \([-1, -1]\):
\begin{itemize}
    \item Class 1 (\( y = 1 \)): \([1, 1]\), \([-1, -1]\)
    \item Class -1 (\( y = -1 \)): \([1, -1]\), \([-1, 1]\).
\end{itemize}
Due to the non-separability, the initial fitting plane cannot perfectly classify the two classes and can only reduce the MSE error value to prevent it from being too large.

This pre-training also uses the \(n = 100\) samples generated from \(P_{S_0}\). In this case, the prior has the same form as in equation (\ref{equ:prior_n}).

\subsubsection{Worst-Case Distribution Impact}

\begin{figure}[t!]
    \centering
    \includegraphics[width=0.48\textwidth]{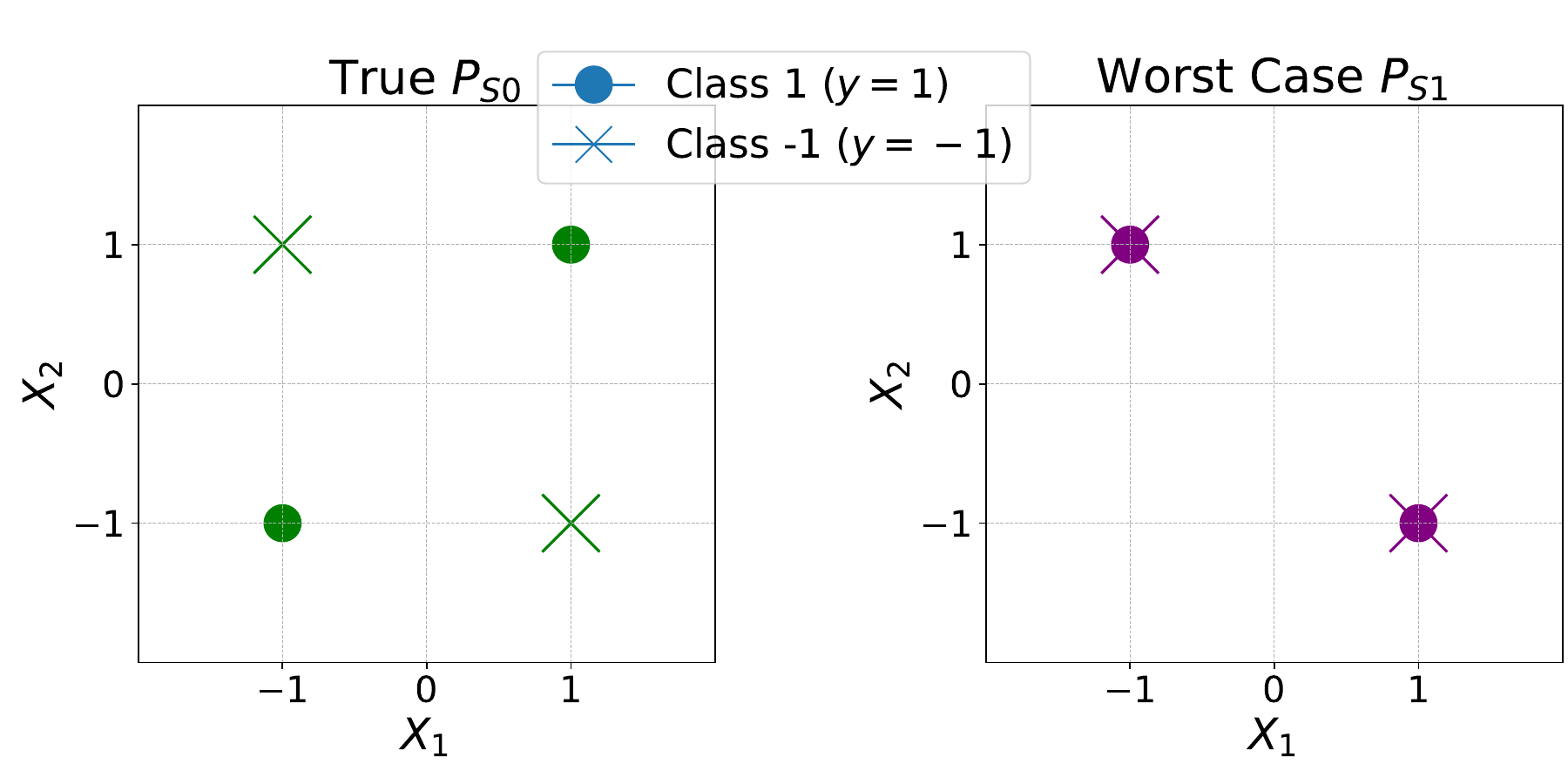}
    \caption{Linearly Non-separable Data Points (Case 2). The initial data distribution \(P_{S_0}\) is represented on the left. Under the worst-case data distribution \(P_{S_1}\) on the right, the class labels have shifted.}\label{Fig: LNS_case}
\end{figure}

Using Algorithm~\ref{alg:RefinementPx}, we identify the worst-case data distribution for the \((n, \pi_n(w), L_E(w, Z_{n+1}))\)-Gibbs algorithm. Figure~\ref{Fig: LNS_case} visualizes the worst-case data distribution \(P_{S_1}\). While the coordinates remain the same as in \(P_{S_0}\), the target labels are altered, affecting class assignments. In \(P_{S_1}\):
\begin{itemize}
    \item Class 1 (\( y = 1 \)): \([1, -1]\), \([-1, 1]\)
    \item Class -1 (\( y = -1 \)): \([1, 1]\), \([-1, -1]\)
\end{itemize}
% In Figure~\ref{Fig: LNS_case}, circles represent Class 1 (\( y = 1 \)) and crosses represent Class -1 (\( y = -1 \)). The worst-case distribution \(P_{S_1}\) changes the fitting plane, leading to increased misclassification rates and model variance.

This shift in the target outcomes effectively changes the fitting plane. The details of the posterior are available in  Appendix~\ref{app:Gibbs}. Under the worst-case data distribution, a single data point can have two labels with the same probability, making it difficult to distinguish. This ambiguity causes a significant increase in the model's variance, resulting in a more spread-out posterior distribution and increased uncertainty.
    % 

% \todoa{What do you mean by arrows in caption?}

\section{Conclusion and Future Works}

This paper presents a novel approach to computing the capacity of discrete channels in terms of symmetrized Kullback-Leibler (KL) information (\(I_{\mathrm{SKL}}\)). By transforming the problem into a discrete quadratic optimization with simplex constraints, we developed the Max-SKL algorithm. This algorithm symmetrizes the KL divergence matrix and employs iterative updates to ensure non-decreasing objective values while maintaining valid probability distributions. In summary, our work enhances the understanding of channel capacity and the impact of input distributions, providing a valuable tool for both theoretical analysis and practical applications.

% We validated our approach on Binary Symmetric Channels and Binomial Channels, demonstrating strong consistency with theoretical capacities. Additionally, we applied our algorithm to the Gibbs channel, effectively identifying worst-case data distributions that maximize generalization error in learning contexts.

Future research could extend this algorithm to more complex channel models and higher-dimensional data distributions, enhancing its applicability and performance. Specifically, our Max-SKL algorithm, currently designed for discrete inputs, is undergoing further development to accommodate continuous inputs using Random Matrix Theory based on mean-field theory.

% \newpage

\bibliographystyle{ieeetr}
\bibliography{ref}

% \newpage
% \clearpage
\appendices
% \section{Appendix}
\section{Lautum Information is Not Concave for Input Distribution}
\label{app:non-concave}
We start with the original proof in \cite{palomar2008lautum} showing that Lautum information \(L(X; Y)\) is concave with respect to the input distribution \(P_X\). 

\textbf{Original proof:}
The proposition was based on the following formulation:
\begin{align}
L(X; Y) &= D(P_Y \| P_{Y|X} | P_X) \nonumber\\
&= D\left(\int P_{Y|X} dP_X \| P_{Y|X} | P_X\right)
\end{align}

To demonstrate concavity, the proof considered a binary random variable \(U\) with values:
\begin{align}
U = 
\begin{cases}
1, & \text{with probability } \alpha, \\
0, & \text{with probability } 1 - \alpha,
\end{cases}
\end{align}
where \(\alpha \in [0, 1]\) represents the probability that \(U = 1\).

Let \(X_0\) and \(X_1\) be two independent random variables. The mixed distribution is defined as:
\begin{align}
P_{X_U} = (1 - \alpha) P_{X_0} + \alpha P_{X_1}.
\end{align}

Using the Markov chain \( (U, X_0, X_1) - X_U - Y \), the data processing inequality was applied to suggest:
\begin{align}
L(X_U; Y) &\geq L(U, X_0, X_1; Y)\nonumber\\
&= L(X_0, X_1; Y | U) + L(U; Y)\nonumber\\
&\geq L(X_0, X_1; Y | U)\nonumber\\
&= (1 - \alpha) L(X_0; Y) + \alpha L(X_1; Y),
\end{align}
implying that \(L(X; Y)\) is concave in \(P_X\).

\textbf{Critical Review of the Chain Rule:} Upon re-evaluation, we found an issue with the application of the chain rule for mutual information in the context of Lautum information. Specifically, \textbf{the chain rule does not hold for Lautum information.} The detailed steps are as follows:
\begin{align*}
&L(U, X_0, X_1 ; Y) = \mathbb{E}_{P_{U,X_0,X_1}P_Y} \left[ \log \frac{P_{U,X_0,X_1} P_Y}{P_{U,X_0,X_1,Y}} \right],\\
&L(U ; Y) = \mathbb{E}_{P_U P_Y} \left[ \log \frac{P_U P_Y}{P_{U,Y}} \right],\\
&L(X_0, X_1 ; Y | U) = \mathbb{E}_{P_{X_0,X_1}} \mathbb{E}_{P_{Y,U}} \left[ \log \frac{P_{Y,U} P_{X_0,X_1}}{P_{X_0,X_1,Y,U}} \right],\\
&L(U, X_0, X_1 ; Y) - L(U ; Y)\\
&= \mathbb{E}_{P_{X_0,X_1}} \mathbb{E}_{P_YP_U} \left[ \log \frac{P_{X_0,X_1} P_{Y,U}}{P_{U,X_0,X_1,Y}} \right] 
\neq L(X_0, X_1 ; Y | U).
\end{align*}
This discrepancy shows that the concavity argument based on the chain rule is incorrect. Therefore, Lautum information is not concave with respect to the input distribution \(P_X\).

\section{Proof of Proposition \ref{prop:SKL}}\label{app:SKL}
For discrete channel $P_{Y|X}$, the symmetrized KL divergence can be written as 
    % the following expression, as shown in \cite{aminian2015capacity}:
\begin{align}\label{equ:SKL1}
I_{\mathrm{SKL}}(X;Y) &= \sum_{x, y} p(x, y) \log \left( \frac{p(x, y)}{p(x)p(y)} \right) \nonumber \\
&\quad + \sum_{x, y} p(x)p(y) \log \left( \frac{p(x)p(y)}{p(x, y)} \right) \nonumber\\
&=\sum_{x,y} p(x) p(y|x) \log \left( p(y|x) \right)  \\
&\quad - \sum_{x,\tilde{x},y} p(x) p(\tilde{x}) p(y|\tilde{x}) \log \left( p(y|x) \right). \nonumber 
% &= \sum_{x, \tilde{X}, y} p(x)p(\tilde{X})p(y|x) \log \left( \frac{p(y|x)}{p(y|\tilde{X})} \right).
\end{align}
Since \(X\) and \(\tilde{X}\) are independent copies with the same distribution, we can interchange \(x\) and \(\tilde{x}\) in~\eqref{equ:SKL1}. This allows us to rewrite the expression as:
\begin{align}
&\sum_{x,\tilde{x},y} p(x) p(\tilde{x}) p(y|\tilde{x}) \log \left( p(y|x) \right) \nonumber \\
&=  \sum_{x,\tilde{x},y} p(x) p(\tilde{x}) p(y|x) \log \left( p(y|\tilde{x}) \right).
\end{align}

Therefore, we have
\begin{align}
& I_{\mathrm{SKL}}(X;Y) =\sum_{x,\tilde{x},y} p(x)p(\tilde{x}) p(y|x) \log \left( p(y|x) \right) \nonumber\\
&\qquad- \sum_{x,\tilde{x},y} p(x) p(\tilde{x}) p(y|x) \log \left( p(y|\tilde{x}) \right) \\
&= \sum_{x,\tilde{x},y} p(x) p(\tilde{x}) \left[ p(y|x) \log \left( p(y|x) \right) - p(y|x) \log \left( p(y|\tilde{x}) \right) \right] \nonumber\\
% &= \sum_{x,\tilde{x},y} p(x) p(\tilde{x}) p(y|x) \left[ \log \left( p(y|x) \right) - \log \left( p(y|\tilde{x}) \right) \right] \\
&= \sum_{x,\tilde{x},y} p(x) p(\tilde{x}) p(y|x) \log \left( \frac{p(y|x)}{p(y|\tilde{x})} \right).\nonumber
% & = \sum_{x,\tilde{x},y} p(x) p(\tilde{x}) D\left( P_{Y|X=x} \| P_{Y|X=\tilde{x}} \right)
% .
\end{align}

We notice that if we sum over \( y \), the inner term can be written as the KL divergence between the channel conditional on $x$ and its independent copy $\tilde{x}$, which completes the proof. We note that a similar treatment of $I_{\mathrm{SKL}}$ can be found in the notes \cite{Polyanskiy_Wu_2024}.

\section{Binary Asymmetric Channel}\label{App:BAC}

To further validate our algorithm, we conducted experiments on a Binary Asymmetric Channel (BAC) with the following crossover probabilities $p=0.1$ and $q=0.6$.
% \begin{itemize}
%     \item \( p = 0.1 \)
%     \item \( q = 0.6 \)
% \end{itemize}
Then, the transition probabilities for the BAC are given by:
\begin{equation}
    P_{Y|X}(y|x) = \begin{pmatrix}
1 - p & p \\
q & 1 - q
\end{pmatrix}.
\end{equation}

Our goal is to demonstrate that the caid of mutual information \( I(X;Y) \) capacity and that of the symmetric KL \( I_{\mathrm{SKL}} \) capacity can be different, as well as to validate that our algorithm achieves the symmetric KL capacity and provides the correct caid.

\begin{figure}[h]
    \centering
    \includegraphics[width=0.42\textwidth]{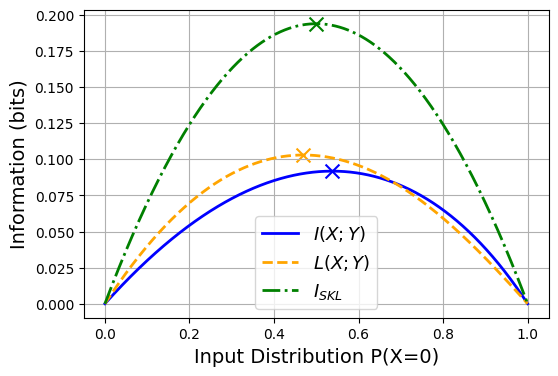}
    \caption{Binary Asymmetric Channel Experiment. The plot shows mutual information \( I(X;Y) \), Lautum information \( L(X;Y) \), and their sum \( I_{\mathrm{SKL}} \). The crosses indicate the input distributions that achieve the respective capacities.}\label{fig:BAC}
\end{figure}

Figure~\ref{fig:BAC} demonstrates that the input distribution that maximizes mutual information is different from the one that maximizes the symmetrized KL information. This highlights the difference in optimization targets between mutual information and symmetrized KL information. 

\begin{figure}[h]
    \centering
    \includegraphics[width=0.42\textwidth]{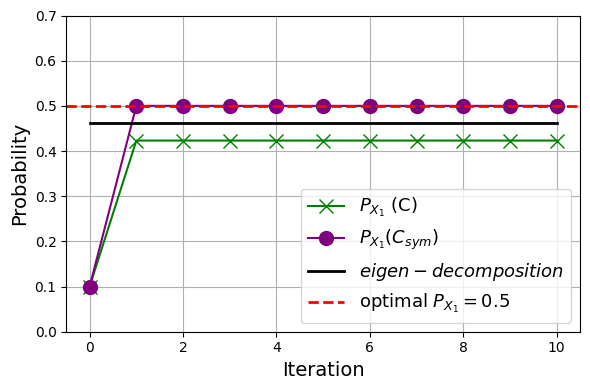}
    \caption{This figure shows the objective input maximizing \(I_{\mathrm{SKL}}\) is 0.5 for the Binary Asymmetric Channel (BAC), indicated by the dashed line. Our algorithm consistently reaches the objective input, whereas, without the symmetrized version, the target capacity cannot be reached. The eigendecomposition method, although closer, also fails to converge to the theoretical value, underscoring the effectiveness of our algorithm.}
\label{fig:BAC_iter}
\end{figure}

Figure~\ref{fig:BAC_iter} shows the convergence of our algorithm towards the symmetric KL capacity. The optimized input distribution obtained from our algorithm is \( P(X=0) = 0.5 \), which is consistent with the theoretical prediction. The Figure~\ref{fig:BAC_iter} includes a comparison with the eigendecomposition method, where the eigenvector corresponding to the largest eigenvalue is normalized to form a distribution. Both the non-symmetrized approach and the eigendecomposition method fail to converge to the theoretical value, highlighting the robustness and accuracy of our algorithm.

% Interestingly, the input distribution that achieves the symmetric KL capacity in the BAC setting is uniform (\( P(X=0) = 0.5 \)), similar to the BSC case. This observation might be due to the symmetric nature of the optimization problem and the fact that the diagonal elements of the quadratic optimization matrix are zero. This leads to an equal weighting of the input probabilities in the optimal solution.

% \bu{This is not the explanation for this observation.}

\newpage
\section{Other details for the Binomial Channel}\label{Bino}
The channel distribution for the binomial channel setting with n=10:
\begin{figure}[H]
    \centering
    \includegraphics[width=0.5\textwidth]{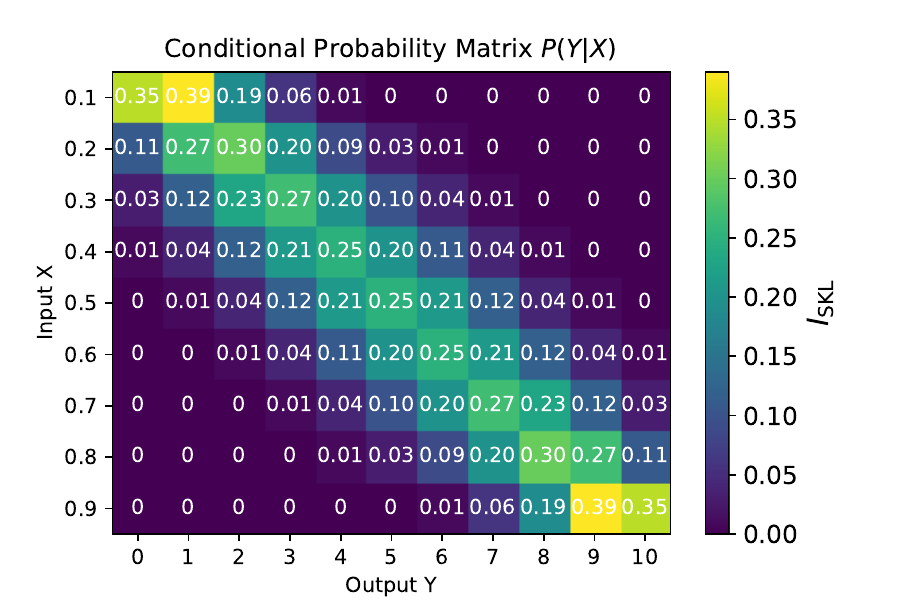}
    \caption{Conditional Probability Matrix \(P(Y|X)\) for the Binomial Channel(n=10).}
    \label{fig:channel_matrix}
\end{figure}

When \(n=100\), the distinguishability between input distributions is less evident, leading to a more distributed optimal input distribution for achieving capacity.

Optimal Input Distribution for \(I_{\mathrm{SKL}}\):
Using the Max-SKL algorithm, the capacity-achieving input distribution (acid) is:
\[
P(X) = [0.3, 0, 0.03, 0.17, 0, 0.17, 0.03, 0 0.3 ]
\]

\begin{figure}[H]
    \centering
    \includegraphics[width=0.5\textwidth]{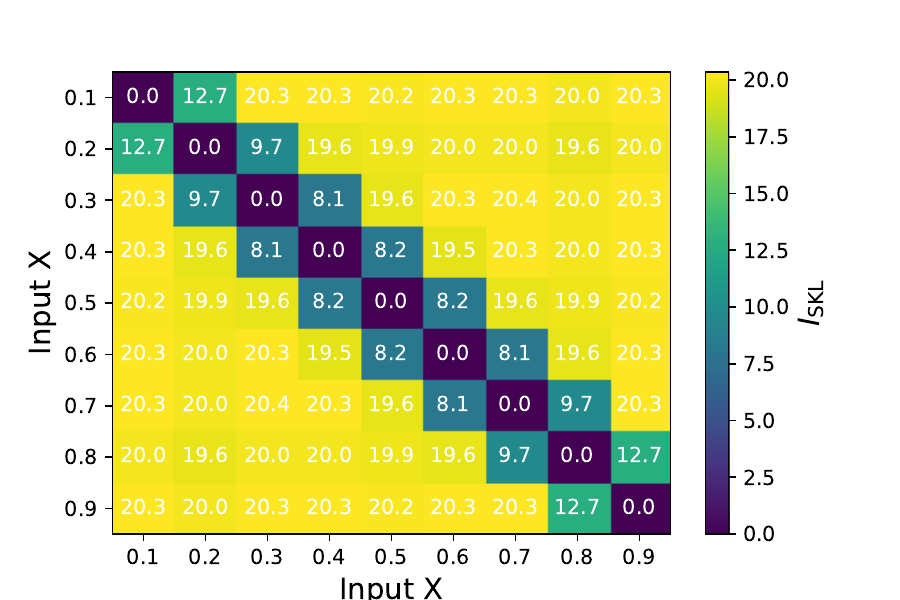}
    \caption{Matrix of KL Divergence \(C_{sym}\) for \(n=100\). The distinguishability is less pronounced, resulting in a more evenly spread distribution to achieve capacity.}
    \label{fig:kl_divergence_n100}
\end{figure}

\section{Experiment detail}
\label{app:Gibbs}
This appendix provides detailed numerical results for the posterior values and probability changes in data distributions across iterations for both Case 1 (Linearly Separable Data Points) and Case 2 (Linearly Nonseparable Data Points). These results support the analysis presented in the main body of the paper and offer a deeper understanding of how the worst-case data distribution affects the classifier's performance.

\subsubsection{Case 1: Linear Separable Data Points}

Table~\ref{tab:case1_iter} shows the changes in the probability distribution of the data points across iterations. Initially, the probabilities are set to 0.25 for each point. In the worst-case scenarios, these probabilities are adjusted to degrade the classifier's performance.

\begin{table}[h] 
    \centering
    \caption{Probability Changes in Data Distribution Across Iterations for Case 1}\label{tab:case1_iter}
    \begin{tabular}{cccccc}
        \toprule
        \textbf{(X, Y)} & \textbf{True \(P_{S_0}\)} & \textbf{Worst Case \(P_{S_1}\)} & \textbf{Worst Case \(P_{S2}\)} \\
        \midrule
        $([1, 1], 1)$ & $0.25$ & $0$ & $0.25$ \\
        $([1, -1], -1)$ & $0.25$ & $0$ & $0.25$ \\
        $([-1, 1], 1)$ & $0.25$ & $0$ & $0.25$ \\
        $([-1, -1], -1)$ & $0.25$ & $0$ & $0.25$ \\
        $([1, 1], -1)$ & $0$ & $0.25$ & $0$ \\
        $([1, -1], 1)$ & $0$ & $0.25$ & $0$ \\
        $([-1, 1], -1)$ & $0$ & $0.25$ & $0$ \\
        $([-1, -1], 1)$ & $0$ & $0.25$ & $0$ \\
        \bottomrule
    \end{tabular}
\end{table}

Table~\ref{tab:case1_pos} provides the posterior means and covariances for the initial and worst-case scenarios. These values indicate how the classifier's decision boundary shifts under the worst-case distribution, leading to degraded performance.

\begin{table}[ht]
    \centering
    \caption{Posterior Mean and Covariance for Case 1} \label{tab:case1_pos}
    \begin{tabular}{cc}
        \toprule
        \textbf{Posterior} & \textbf{Value} \\
        \midrule
        \(\mu_{\text{initial}}\) & $\begin{bmatrix} 0 \\ 9.89999000 \times 10^{-1} \end{bmatrix}$ \\
        \(\Sigma_{\text{initial}}\) & $\begin{bmatrix} 0.010001 & -0.00010001 \\ -0.00010001 & 0.010001 \end{bmatrix}$ \\
        \(\mu_{\text{worst}}\) & $\begin{bmatrix} -1.0001 \times 10^{-4} \\ -9.89999 \times 10^{-1} \end{bmatrix}$ \\
        \(\Sigma_{\text{worst}}\) & $\begin{bmatrix} 0.01 & -0.00 \\ -0.00 & 0.01 \end{bmatrix}$ \\
        \bottomrule
    \end{tabular}
\end{table}
\begin{figure}[h!]
    \centering
    \includegraphics[width=0.42\textwidth]{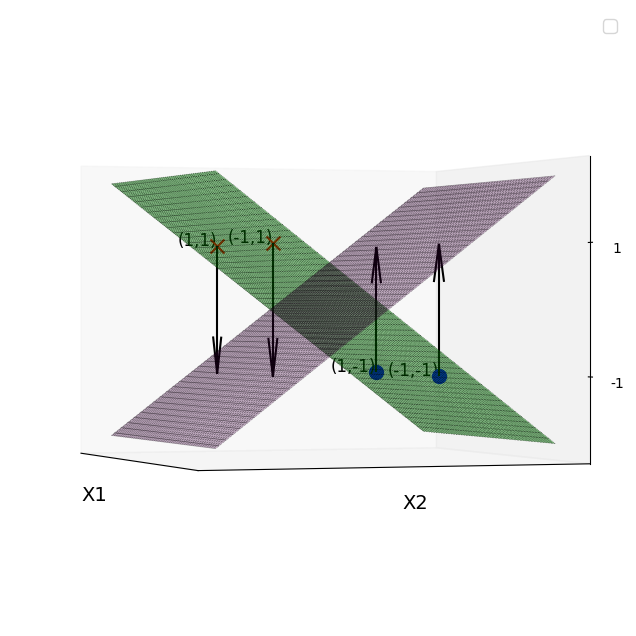}
    \caption{Case 1: Linear Separable Data Points. The initial classifier, represented by the green plane \( y = x_2 \), correctly classifies the data points. Under the worst-case data distribution, the classifier's decision boundary shifts to the purple plane \( y = -x_2 \), leading to misclassification. The arrows indicate the shift in classification from the initial data distribution to the worst-case distribution.}

\end{figure}

% \bu{Move this to appendix}
Interestingly, if we continue to iterate to find the worst case $P_{S_2}$ of the worst data $P_{S_1}$ from the first iteration, the class assignments revert to the initial distribution:
\begin{itemize}
    \item Class 1 (\( y = 1 \)): \([1, 1]\), \([-1, -1]\)
    \item Class -1 (\( y = -1 \)): \([1, -1]\), \([-1, 1]\)
\end{itemize}

This iterative process causes the fitting plane to revert to the initial decision boundary \( y = x_2 \), which correctly classifies the initial data but misclassifies the worst-case data.
\subsubsection{Case 2: Linearly Nonseparable Data Points}

Table~\ref{tab:case2_iter} shows the changes in the probability distribution of the data points across iterations for Case 2. These adjustments highlight how the worst-case distribution alters the data to increase the classifier's uncertainty.
\begin{table}[h!]
    \centering    \caption{Probability Changes in Data Distribution Across Iterations for Case 2}\label{tab:case2_iter}
    \begin{tabular}{cccc}
        \toprule
        \textbf{(X, Y)} & \textbf{True \(P_{S_0}\)} & \textbf{Worst Case \(P_{S_1}\)} & \textbf{Worst Case \(P_{S2}\)} \\
        \midrule
        $([1, 1], 1)$ & $0.25$ & $0$ & $0.25$ \\
        $([1, -1], -1)$ & $0.25$ & $0.25$ & $0$ \\
        $([-1, 1], -1)$ & $0.25$ & $0.25$ & $0$ \\
        $([-1, -1], 1)$ & $0.25$ & $0$ & $0.25$ \\
        $([1, 1], -1)$ & $0$ & $0$ & $0.25$ \\
        $([1, -1], 1)$ & $0$ & $0.25$ & $0$ \\
        $([-1, 1], 1)$ & $0$ & $0.25$ & $0$ \\
        $([-1, -1], -1)$ & $0$ & $0$ & $0.25$ \\
        \bottomrule
    \end{tabular}
\end{table}

Table~\ref{tab:case2_pos} provides the posterior means and covariances for the initial and worst-case scenarios in Case 2. These values reflect how the classifier's uncertainty increases under the worst-case distribution, as indicated by the broader posterior covariance.

\begin{table}[h!]
    \centering
    \caption{Posterior Mean and Covariance for Case 2}\label{tab:case2_pos}
    \begin{tabular}{cc}
        \toprule
        \textbf{Posterior} & \textbf{Value} \\
        \midrule
        \(\mu_{\text{initial}}\) & $\begin{bmatrix} 0.0497 \\ 0.0295 \end{bmatrix}$ \\
        \(\Sigma_{\text{initial}}\) & $\begin{bmatrix} 0.0100 & -0.0001 \\ -0.0001 & 0.0100 \end{bmatrix}$ \\
        \(\mu_{\text{worst}}\) & $\begin{bmatrix} -0.0050 \\ 0.0050 \end{bmatrix}$ \\
        \(\Sigma_{\text{worst}}\) & $\begin{bmatrix} 0.5025 & 0.4975 \\ 0.4975 & 0.5025 \end{bmatrix}$ \\
        \bottomrule
    \end{tabular}
\end{table}

By analyzing these cases, we illustrate how the worst-case data distribution can significantly impact the performance of a classifier. In the linearly separable case, the worst-case distribution inverts the classifier's decision boundary, while in the linearly nonseparable case, it increases the classifier's uncertainty by broadening the posterior distribution. These experiments highlight the importance of understanding and mitigating the effects of adversarial data distributions in machine learning.

\begin{figure}[H]
    \centering
    \includegraphics[width=0.42\textwidth]{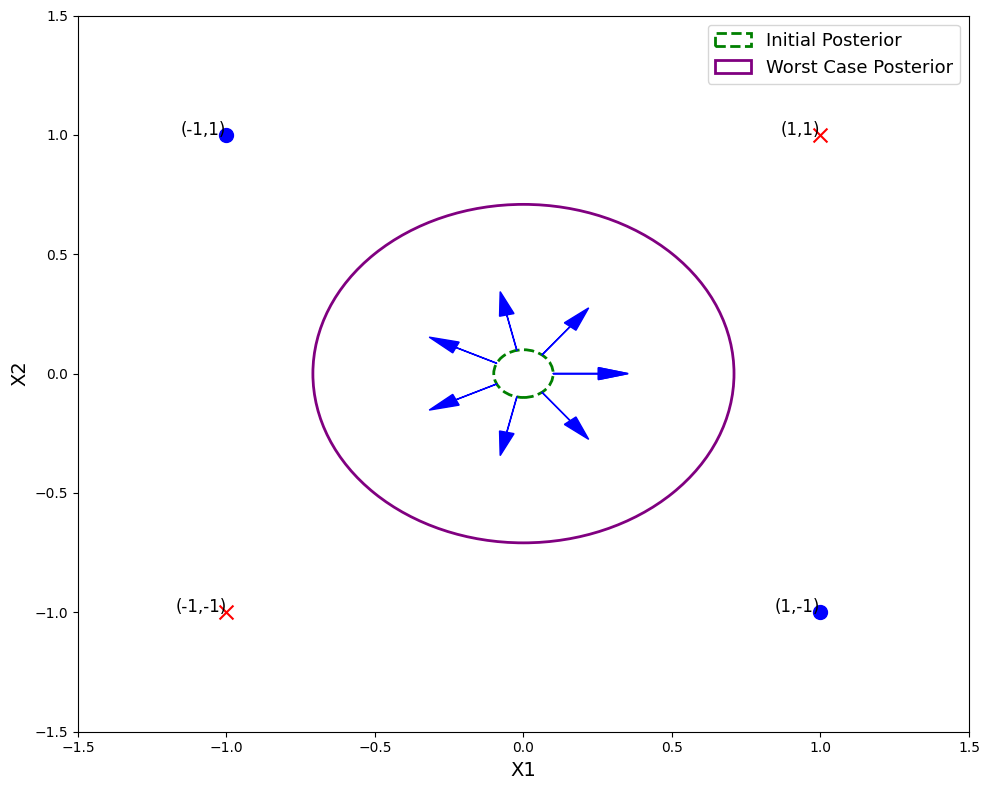}
    \caption{Case 2: Linearly Nonseparable Data Points. The initial classifier, represented by the green ellipse, shows the posterior distribution based on 100 samples as prior. Under the worst-case data distribution, the classifier's variance increases significantly, represented by the red ellipse, indicating a broader posterior distribution and increased uncertainty. The arrows indicate the direction of increased uncertainty from the initial data distribution to the worst-case distribution.}
\end{figure}

\end{document}